\documentclass[aps,prb,twocolumn,letterpaper,superscriptaddress,showpacs]{revtex4}
\usepackage{CJK}
\usepackage{keyval}%
\usepackage{graphicx}
\usepackage{dcolumn}
\usepackage{bm}
\usepackage{color}
\usepackage{amsmath}

\setkeys{Gin}{width=0.8\columnwidth,clip=true}
\newcommand{\ignore}[1]{}

\newcommand{\ket}[1]{|{#1}\rangle}
\newcommand{\bra}[1]{\langle{#1}|}

\begin{document}
\begin{CJK*}{UTF8}{bsmi}
\title{Unfolding method for the first-principles LCAO electronic structure calculations}
\author{Chi-Cheng Lee (%
李啟正
)}
\affiliation{School of Materials Science, Japan Advanced Institute of Science and Technology (JAIST),
1-1 Asahidai, Nomi, Ishikawa 923-1292, Japan}%
\author{Yukiko Yamada-Takamura
}
\affiliation{School of Materials Science, Japan Advanced Institute of Science and Technology (JAIST),
1-1 Asahidai, Nomi, Ishikawa 923-1292, Japan}%
\author{Taisuke Ozaki
}
\affiliation{School of Materials Science, Japan Advanced Institute of Science and Technology (JAIST),
1-1 Asahidai, Nomi, Ishikawa 923-1292, Japan}%
\affiliation{Research Center for Simulation Science, Japan Advanced Institute of Science and Technology (JAIST), 
1-1 Asahidai, Nomi, Ishikawa 923-1292, Japan}

\date{\today}

\begin{abstract}
Unfolding the band structure of a supercell to a normal cell enables us to investigate how 
symmetry breakers such as surfaces and impurities perturb the band structure of the normal cell.
We generalize the unfolding method, originally developed based on Wannier functions, to the linear combination 
of atomic orbitals (LCAO) method, and present a general formula to calculate the unfolded spectral weight.
The LCAO basis set is ideal for the unfolding method because 
of the invariance that basis functions allocated to each atomic species are invariant regardless of 
existence of surface and impurity. The unfolded spectral weight is well defined by the property of the 
LCAO basis functions. In exchange for the property, the non-orthogonality of the 
LCAO basis functions has to be taken into account. We show how the non-orthogonality can be properly 
incorporated in the general formula. As an illustration of the method, we calculate the dispersive 
quantized spectral weight of ZrB$_2$ slab and show strong spectral broadening in the out-of-plane direction, 
demonstrating the usefulness of the unfolding method.
\end{abstract}


\maketitle
\end{CJK*}
\section{Introduction}

The Kohn-Sham (KS) framework \cite{Kohn} within the density functional theory (DFT) allows us to investigate 
a wide variety of imperfect materials such as surface, impurities, and vacancies.\cite{Ashcroft, Martin}
A widely used method to perform first-principles calculations of such systems is to introduce a supercell 
which makes studies of various forms of imperfections possible.\cite{Ashcroft, Martin}
Not restricted by the periodic boundary condition, the Bloch theorem can also be applied to study a non-periodic system 
by introducing a large supercell which simulates a system where the translational symmetry is highly broken, 
e.g., the presence of a surface.\cite{Payne} 
However, there are at least two drawbacks introduced by the large supercell in analyzing 
the electronic structure. First, the bands folded heavily in the small first Brillouin zone (BZ) corresponding to 
the large supercell makes it difficult to analyze how symmetry breakers such as surfaces and impurities 
perturb the band structure of the normal cell, where by the {\it normal cell} we mean a unit cell which is smaller 
than the supercell, less imperfect, and gives a reference of band structure.
Second, it would be difficult to directly compare the heavily folded bands with experimental results. 
For example, the band structure calculated for the supercell cannot be directly compared with the spectra measured by 
the angle resolved photoemission spectroscopy (ARPES) without further considering the proper spectral weight, 
the imaginary part of the retarded one-particle Green function.\cite{Faulkner,Onida} 

It would be desirable to develop a method which represents the band structure or spectral function 
of the supercell in terms of eigenstates of a chosen normal cell in order to relieve the two drawbacks in the 
supercell calculations. Unfolding methods have been proposed as an idea of realizing the change of 
representation of spectral function in terms of the eigenstates of the normal cell, and implemented 
in a wide variety of ways. 
For example, one can efficiently and rigorously calculate the band structure of the normal cell 
from a $\Gamma$-point calculation of the large supercell without imperfection 
by using maximally localized Wannier functions (WFs).\cite{Marzari} 
Another exact method for unfolding the band structure of a perfect supercell into a bulk dispersion relation, which 
can be compared to experiments, has been proposed in tight-binding calculations, and was further applied to an imperfect
supercell via an averaged Hamiltonian.\cite{Boykin} 
Unfolding methods have also been introduced to the plane-wave basis sets.\cite{Voicu,Zunger} Recently, a new unfolding 
approach by using energy-resolved symmetry-respecting WFs has been proposed by representing the 
spectral function from the supercell calculation in the basis of a conceptual normal 
cell instead of the representation by the eigenstates of the supercell.\cite{Wei} 
The approach allows us to uniquely determine the unfolded spectral weight for a chosen normal cell 
via WFs and the geometrical structure of the supercell. 

While the unfolding methods based on the WFs have been successful in providing detailed physical 
insights on various systems,\cite{Marzari,Wei,Tom,Arita,Kang} the WFs need to be constructed
in the actual implementation.\cite{Vanderbilt, Scalettar}
For large-scale systems the construction of the WFs can be time-consuming, 
which may hamper the applicability of the unfolding methods to large-scale systems. 
In addition to this, one may encounter a difficulty that one to one correspondence between 
WFs defined in the supercell and normal cell could not be well defined owing to the gauge freedom\cite{Vanderbilt} 
during the construction of the Wannier functions. 
Since the most essential step in the unfolding method based on the WFs is to determine 
one to one correspondence between WFs defined in the supercell and normal cell, 
the gauge freedom has to be utilized so that the symmetry and shape of WFs can be approximately 
identical between the two cells.\cite{Wei} The second issue may also pose a difficulty in applying 
the unfolding method to systems with strong perturbation.
Thus, it would be physically more preferable if the localized basis functions are identical for the same 
atomic species in all the normal cells arranged in the supercell by following the proposed unfolding 
method.\cite{Wei} For the unfolding method, the linear combination of atomic 
orbitals (LCAO) method\cite{Koepernik,Blum,Soler,Frisch,Andersen,Ozaki} 
can be regarded as an ideal framework in which the same basis functions are allocated for each atomic 
species regardless of existence of surface and impurity. It is apparent that one can easily establish 
one to one correspondence between AOs located in the supercell and normal cell without any ambiguity.
In this paper we extend the unfolding method,\cite{Wei} originally developed based on Wannier functions, to the LCAO method, 
and present a general formula to calculate the unfolded spectral weight for a chosen normal cell.
In exchange for the invariance of the LCAO basis functions, the non-orthogonality between the LCAO basis functions 
has to be taken into account. We show how the non-orthogonality can be properly incorporated in the general formula.
In addition to the ideal property of the LCAO method to the unfolding method, efficient computational methods have 
been developed by making use of the locality of the LCAO basis functions. 
Thus, it can be expected that the generalization to the LCAO method extends the applicability of the unfolding 
method to large-scale systems.\cite{Yang,Koepernik,Blum,Soler,Frisch,Andersen,Ozaki2,Ozaki3}

As an illustration of the method, we apply the method for recovering the bulk dispersion relation in the out-of-plane 
direction from a DFT electronic structure calculation for a ZrB$_2$ slab with the (0001) surface, where only 
one $k$-point sampling is needed along the direction. 
A graphene counterpart, silicene, has recently been epitaxially grown on the ZrB$_2$(0001) surface.\cite{Fleurence} 
Therefore, it would be important to investigate the surface and slab states of ZrB$_2$ in order to deeply understand 
the newfound silicene which could be constituent of future devices as well as extensively studied 
graphenes.\cite{Ohta,Castro,Fleurence,vogt} 
It is shown that the unfolding method reveals the dispersive quantized spectral weight of the ZrB$_2$ slab
and strong spectral broadening in the out-of-plane direction, which is expected to be measured by experiments.

The paper is organized as follows: In Sec.~II, the concept of the unfolding method is introduced. 
After introducing the concept, the unfolding formula is derived in Sec.~III. In Sec.~IV, we illustrate 
how to calculate the unfolded spectral weight along the out-of-plane direction in the bulk BZ from a slab 
calculation for ZrB$_2$ where a dispersive quantized spectral weight with strong broadening can be revealed.
Finally, we summarize our study in Sec.~V.

\section{The concept of unfolding band structures}

\begin{figure}[tbp]
\includegraphics*[width=1.00\columnwidth,angle=0]{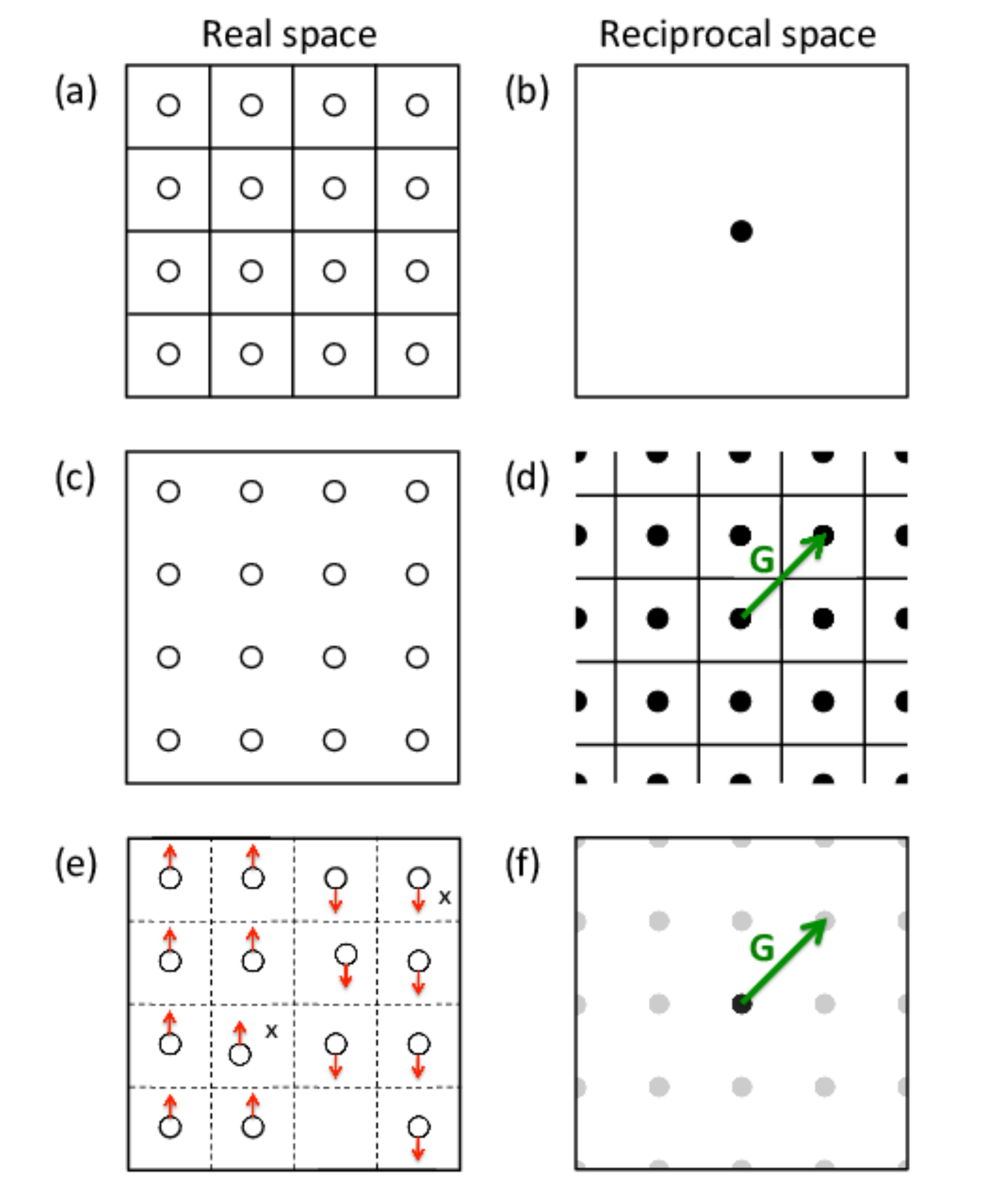}
\caption{\label{fig:fig1}
(a) A two-dimensional model system which contains one atom per unit cell (open circles).
(b) The model system is assumed to have an eigenstate at the Fermi point which is located at the $\Gamma$ point. 
The boundary of the first BZ is represented by the square.
The eigenstate at the Fermi point and the intensity of the spectral weight, which is exactly one, are 
represented by the black solid circle.
(c) The perfect 4$\times$4 supercell. (d) The Fermi points for the perfect 4$\times$4 supercell.
The zones are periodic and the spectral weight is the same upon arbitrary integer shifts of reciprocal lattice 
vectors ($G$). (e) Several imperfections can be introduced to the supercell, for example, displacements, 
vacancies, impurities (plotted by x), and magnetic orders (red arrows). 
(f) The expected proper spectral weight with a weak translational symmetry breaking corresponding to Fig.~\ref{fig:fig1}(e).
The gray solid circle indicates a relatively low intensity of the spectral weight.
}
\end{figure}

The concept of the unfolding method is introduced in this section. The physical meaning of the conceptual normal 
cell is also discussed here, which is the key element for the unfolding method. 
With the introduction of the conceptual normal cell, the method can uniquely determine the spectral weight 
in the BZ of the conceptual normal cell, which directly presents the strength of the translational symmetry 
breaking to the band structure of the normal cell. 
The concept is illustrated with a two-dimensional model system of which primitive cell contains one atom per 
unit cell as shown in Fig.~\ref{fig:fig1} (a). It is assumed that the Fermi point is located at the $\Gamma$ 
point and only a single eigenstate, whose spectral weight is exactly one, is involved at the Fermi point as 
shown in Fig.~\ref{fig:fig1} (b), where we used the Fermi {\it point} instead of the Fermi {\it surface} due to 
the involvement of only the $\Gamma$ point. 
To simplify the illustration, we focus only on the spectral weight on the Fermi 
point, but the idea can be applied for all the $k$-dependent eigenstates. 

To study the same system, one can perform a perfect supercell calculation, for example, the 4$\times$4 supercell 
as shown in Fig.~\ref{fig:fig1} (c). As long as there is no additional symmetry breaking, the physical properties 
should be exactly the same as those of the primitive cell. Therefore, the primitive unit cell is referred to as 
the normal cell. 
However, it is confirmed that the Fermi points obtained by the supercell calculation as shown 
in Fig.~\ref{fig:fig1} (d) are different from those of the normal cell.
Because the first BZ of the supercell is smaller and is translationally symmetric upon arbitrary integer shifts 
of reciprocal lattice vectors ($G$), the repeated counterparts make the comparison difficult.
In order to have a direct comparison between these two calculations, one needs to unfold the band structure 
from the first BZ of the supercell (the reduced-zone representation) to the BZ of the normal cell.
In other words, we would like to recover a proper spectral weight from the periodic-zone representation 
to the extended-zone representation. Obviously, the unfolded band structure is invariant against any arbitrary 
choice of the supercell without perturbation . 
In this case, the unfolded band structure should exhibit the same Fermi point as shown in Fig.~\ref{fig:fig1} (b). 

In case that the system undergoes translational symmetry breaking via the imperfections as shown in Fig.~\ref{fig:fig1} (e), 
it is apparent that first-principles calculations have to be performed for the supercell instead of a smaller unit cell. 
However, it would be expected that a similar spectral weight to Fig.~\ref{fig:fig1} (b) as illustrated in Fig.~\ref{fig:fig1} (f) is recovered 
by introducing the normal cell if the symmetry breaking is not strong just like the case of the 4$\times$4 perfect supercell.
It is also clear that the spectral weight of the supercell eigenstate (exactly one) may not be directly used to 
compare with ARPES. The measured weight cannot be suddenly modified by the introduction of a weak perturbation. 
That is, it is possible that the periodicity of the supercell BZ is not the periodicity of the observed spectral weight. 
Obviously, how to obtain the proper spectral weight via unfolding is an important issue.
A solution to do that is simple at least conceptually. We only have to represent the spectral function obtained by 
the supercell calculation in the eigenstates of the normal cell. 
However, the definition of the normal cell may not be so obvious in general, since the translational symmetry is 
broken in the supercell. 
In the case as shown in Fig.~\ref{fig:fig1} (e), a rather obvious choice is the unit cell plotted with the dashed lines 
as the same as the primitive unit cell as shown in Fig.~\ref{fig:fig1} (a). However, that is not the only choice. 
In fact, the normal cell in the unfolding method is regarded as a conceptual unit cell which defines
one to one correspondence between localized functions such as WFs and AOs in the supercell and normal cell.
If a proper unit cell corresponding to the periodicity of the observed spectral weight is chosen as the normal cell,
the resultant spectral weight in the corresponding BZ can be compared to ARPES after the polarization dependent 
dipole matrix element is taken into account within the sudden approximation.\cite{Wei} 
If a reference unit cell is chosen as the normal cell, the resultant spectral weight will provide information of 
the translational symmetry breaking to the reference system. Therefore, it would be anticipated that the unfolding 
method can be utilized for a wide variety of studies.

In the current approach, once a unit cell is chosen as the normal cell, we only have to consider one to one 
correspondence between AOs in the supercell and normal cell in representing the spectral function of the supercell 
in the eigenstates of the normal cell. Such an assignment is straightforwardly performed in the LCAO method
since the same LCAO basis functions are allocated to each atomic species. 
Then, the information of the translational symmetry breaking in the supercell is stored in two aspects. 
One is that the symmetry breaking is recorded in the LCAO coefficients of the supercell eigenstates. 
The other is that the symmetry breaking is built in the position of the basis functions in the supercell. 
The latter is taken into account by overlap integrals between the LCAO basis functions.
Since the eigenstates of the supercell and the overlap integrals are obtained by a conventional first-principles 
calculation, we discuss in Sec.~III how the proper spectral weight can be calculated from the 
two quantities in detail.

\section{The unfolding formula}

In the section, we present a general formula to calculate the unfolded spectral weight. 
As discussed in the previous section, the proper spectral weight defined in the BZ of a chosen conceptual normal cell
is evaluated by using the eigenstates of the supercell and the overlap matrix elements of the LCAO basis functions. 
The strength of each band's coupling to the symmetry breaker (e.g., impurities, vacancies, dopants, and lattice distortions) 
can be observed via the unfolded spectral weight. 

Let us start to introduce the spectral function $\hat{A}(\omega)$ defined as the imaginary part of one-particle 
Kohn-Sham Green's function: 
\begin{equation}
\label{eq:eqn1}
\begin{aligned}
& \hat{A}(\omega)=- \frac{1}{\pi} {\rm Im} \hat{G}(\omega+i0^+) 
\end{aligned}
\end{equation}
with
\begin{equation}
\label{eq:eqn1-2}
\begin{aligned}
  \hat{G}(z)=\sum_{KJ} \frac{\ket{KJ}\bra{KJ}}{z-\varepsilon_{KJ}},
\end{aligned}
\end{equation}
where $0^+$ is a positive infinitesimal, and $z$ an energy in the complex plane ($z=\omega+i\eta$). 
$\ket{KJ}$ denotes the Bloch state with the crystal momentum $K$ and the band index $J$ 
obtained by a DFT calculation for the supercell. 
Though the non-spin polarized case is considered in the paper for simplicity of notation, 
the generalization is straightforward.
In the LCAO method, atomic basis functions $\{\vert RN\rangle\}$ are placed 
in every unit cell specified with a translational lattice vector $R$, where $N$ is 
a symbolic orbital index consisting of atomic position relative to $R$, a multiplicity index for 
radial functions, an angular momentum quantum number, and a magnetic quantum number.
With the idea of the LCAO method, $\ket{KJ}$ is expanded in a form of linear combination 
of atomic basis functions as
\begin{equation}
\label{eq:eqn2}
 \vert KJ \rangle 
  =
   \sum_{N} C^{KJ}_{N} 
   \vert {KN}\rangle
\end{equation}
with the definition:
\begin{equation}
\label{eq:eqn2-2}
 \vert KN \rangle 
  =
 \frac{1}{\sqrt{L}} \sum_{R} e^{iKR} \vert RN\rangle,
\end{equation}
where $L$ is the number of the unit cells introduced in the Born-von Karman boundary condition, 
and $C^{KJ}_{N}$ the LCAO coefficient. 
Although we use the bracket notation to denote the Bloch state $\ket{KJ}$, 
$\vert KN \rangle$ defined by Eq.~(\ref{eq:eqn2-2}), and the atomic basis function $\vert RN\rangle$, 
the distinction among them is made by alphabet, i.e., the first alphabet is $K$ or $R$, and 
the second is $J$ or $N$(or $M$). 
Then, one of them is uniquely distinguished by the combination of two alphabets. 
The LCAO coefficients and the eigenvalue $\varepsilon_{KJ}$ of $\ket{KJ}$ are calculated 
in the KS framework by solving a generalized eigenvalue problem:
\begin{equation}
\label{eq:eqn3}
H^K C^K = \epsilon^K S(K) C^K
\end{equation}
with the Kohn-Sham Hamiltonian $H$ given by
\begin{equation}
\label{eq:eqn4}
H^{K}_{MN} = \sum_{R} e^{iKR} \bra{0M} \hat{H} \ket{RN},
\end{equation}
and the overlap matrix $S(K)$ given by
\begin{eqnarray}
  S_{MN}(K) = \sum_{R} e^{iKR} S_{0M,RN}, 
\label{eq:eqn5}
\end{eqnarray}
where the overlap matrix element $S_{0M,RN}\equiv \langle 0M|RN\rangle$ reflects the strength 
of the non-orthogonality between the LCAO basis functions.
Since the Bloch state $\ket{KJ}$ and the corresponding eigenvalue $\varepsilon_{KJ}$ are obtained 
by the conventional supercell calculation, the spectral function $\hat{A}(\omega)$ is well defined 
in terms of the representation of the supercell. 

We now consider to represent the spectral function in the eigenstates of a normal cell, which 
is what we mean by {\it unfolding}.
As discussed in the Sec.~II, the normal cell can be chosen as either a unit cell corresponding to 
the periodicity of spectral weight observed in experiments or a reference unit cell depending on 
the purpose under consideration. 
The choice of the normal cell is equivalent to the introduction of a conceptual system of which 
periodicity is the same as that of the normal cell. 
If no symmetry breaker is introduced in the supercell, there is no ambiguity for the introduction 
of the conceptual system. Thus, the unfolding is performed in a precise mathematical sense.
For general cases with symmetry breakers, however, such a periodicity of the normal cell is not hold anymore. 
Nevertheless, we assume that such a conceptual system can be defined and 
the corresponding Bloch state $\ket{kj}$ can be given by  
\begin{equation}
\label{eq:eqn5-2}
 \vert kj \rangle 
  =
   \sum_{n} C^{kj}_{n} 
   \vert {kn}\rangle,
\end{equation}
where $\vert {kn}\rangle$ is the counterpart to Eq.~(\ref{eq:eqn2-2}). 
In the following discussion, uppercase and lowercase letters will be used for indices 
associated with the supercell and the normal cell, respectively. 
The usage of alphabets for the bracket notation in the normal cell follows that of 
the supercell. The substance of the conceptual system that we introduced here might be 
regarded as an average of all the normal cells in the supercell. 
Once the assumption is accepted, the spectral function can be expressed in the representation of the Bloch 
state $\ket{kj}$ of the conceptual system as shown below. By noting that the closure relation
in the non-orthogonal LCAO basis functions is given by 
\begin{equation}
\label{eq:eqn5-3}
 \sum_{kn}\vert \tilde{kn} \rangle \langle kn \vert = \sum_{kn}\vert kn \rangle \langle \tilde{kn} \vert = I
\end{equation}
with the corresponding dual function $\vert \tilde{kn}\rangle$ defined by
\begin{equation}
\label{eq:eqn5-4}
 \vert \tilde{kn} \rangle = \sum_{m} \vert km \rangle S_{mn}^{-1}(k),
\end{equation}
the spectral function $A$ is expressed as 
\begin{equation}
\label{eq:eqn5-5}
  A=\sum_{kj} A_{kj,kj} = \sum_{kn} A_{\tilde{kn},kn},
\end{equation}
where $A_{kj,kj}\equiv \bra{kj} \hat{A} \ket{kj}$ and $A_{\tilde{kn},kn}\equiv \bra{\tilde{kn}} \hat{A} \ket{kn}$.
It is worth mentioning that similar closure relations to Eq.~(\ref{eq:eqn5-3}) are hold for $\vert KN \rangle$, 
$\vert RN\rangle$, $\vert kn \rangle$, and $\vert rn\rangle$. 
By inserting a closure relation $\sum_{KJ} \ket{KJ} \bra{KJ}$ in two places of the last formula of 
Eq.~(\ref{eq:eqn5-5}), and noting that $\hat{A}$ is defined by $\{\ket{KJ}\}$, we obtain 
the following expression for each $k$:
\begin{equation}
\label{eq:eqn5-6}
  A_{kj,kj} = \sum_{mnK} S_{nm}^{-1}(k) \langle km |KJ \rangle A_{KJ,KJ} \langle KJ|kn\rangle,
\end{equation}
where the summation over $j$ and $J$ for the left- and right- hand sides were dropped, respectively. 
The omission of the summations is possible by redefining $j$ so that the state specified with $j$ 
can correspond to the unfolded counterpart to the states specified with $J$. 
The overlap integrals $\langle km |KJ \rangle$ and $\langle KJ\vert kn\rangle$
appearing in Eq.~(\ref{eq:eqn5-6}) bridge two representations of the supercell and the normal cell. 
The integral $\langle km |KJ \rangle$ can be written by using Eq.~(\ref{eq:eqn2}) as 
\begin{equation}
\label{eq:eqn5-7}
  \langle km |KJ \rangle
  = 
   \sum_{M} C^{KJ}_{M}
   \langle km |KM \rangle.
\end{equation}
We further rewrite $\langle km |KM \rangle$ in Eq.~(\ref{eq:eqn5-7}) by making use of two closure relations
$\sum_{r}\vert \tilde{rm} \rangle \langle rm \vert$ and $\sum_{R}\vert RM \rangle \langle \tilde{RM} \vert$ 
as follow:
\begin{eqnarray}
\label{eq:eqn5-8}
  \nonumber
  \langle km |KM \rangle
  &=& 
  \sum_{r,R}
  \langle km | \tilde{rm} \rangle \langle rm 
  \vert RM \rangle \langle \tilde{RM} \vert
  KM \rangle\\
  &=& 
  \sum_{r,R}
  \frac{e^{-ikr}}{\sqrt{l}}
  \langle rm \vert RM \rangle 
  \frac{e^{iKR}}{\sqrt{L}},
\end{eqnarray}
where we used the dual orthonormality relation between the original and its dual functions, 
and $l$ is the number of the unit cells introduced in the Born-von Karman boundary condition
for the normal cell.
As well, it is easy to see that the overlap integral $\langle KJ|kn\rangle$ is written by 
\begin{eqnarray}
\label{eq:eqn5-9}
  \langle KJ|kn\rangle
   = 
  \sum_{N} {C^{KJ}_{N}}^{*}
  \langle KN |kn \rangle
\end{eqnarray}
with 
\begin{eqnarray}
\label{eq:eqn5-10}
  \langle KN |kn \rangle
  =
  \sum_{r',R'}
  \frac{e^{-iKR'}}{\sqrt{L}}
  \langle R'N \vert r'n \rangle 
  \frac{e^{ikr'}}{\sqrt{l}}.
\end{eqnarray}
After we assumed the conceptual system and the Bloch states by Eq.~(\ref{eq:eqn5-2}), 
we have never introduced any approximation. All the derivations up to here are rigorous
in a precise mathematical sense under the assumption of existence of the conceptual system.
However, here we introduce two approximations in evaluating the overlap integrals 
$\langle rm \vert RM \rangle$ and $\langle R'N \vert r'n \rangle$ in Eqs.~(\ref{eq:eqn5-8}) 
and (\ref{eq:eqn5-10}), respectively. 
The first integral $\langle rm \vert RM \rangle$ is evaluated by assuming that 
the position of $\vert RM \rangle$ in real space is the same as that of the counterpart defined 
in the conceptual system, while the second integral $\langle R'N \vert r'n \rangle$ is 
evaluated by assuming that the position of $\vert r'n \rangle$ in real space is the same as 
that of the counterpart defined in the supercell system under investigation. 
The introduction of the approximations may be justified by noting that the true value of each integral 
is approximately given by the mean of the two approximated values if the two approximations are 
independently applied to the single integral. In fact, the {\it conceptual} system can be introduced 
by the approximations without addressing a specific system for the conceptual system as shown later on.
In order to introduce the approximations, we need to establish one to one correspondence between 
AOs in the supercell and normal cell. 
In our case, the establishment of the one to one correspondence is easily realized due to the 
invariance of the LCAO basis functions. 

With the introduction of the two approximations, the second integral $\langle R'N \vert r'n \rangle$
is nothing but the overlap integral used in the supercell calculation. 
Thus, we only have to focus on the first integral $\langle rm \vert RM \rangle$, where $\vert RM \rangle$ 
needs to be relabeled in the representation of the normal cell. By relabeling $R \to R+r_0(M)$ and 
$M\to m'(M)$, we replace as 
\begin{eqnarray}
\label{eq:eqn5-11}
  \langle rm \vert RM \rangle 
  \to 
  \langle rm \vert R+r_0(M), m'(M) \rangle. 
\end{eqnarray}
Although the quantity is nothing but the overlap matrix between AOs in the representation of the normal cell, 
we further rewrite the integral to simplify the last formula of the spectral function $A$ that we have been
pursuing. By inserting the Fourier representation of $\vert rm \rangle$:  
\begin{eqnarray}
\label{eq:eqn5-13}
  \vert rm \rangle &=& \frac{1}{\sqrt{l}}\sum_{k}e^{-ikr}\vert km\rangle,
\end{eqnarray}
and that of $\vert R+r_0(M), m'(M) \rangle$ into Eq.~(\ref{eq:eqn5-11}), and noting that 
\begin{eqnarray}
\label{eq:eqn5-14}
  \nonumber
  \langle km|k' n\rangle  
   &=& 
   \frac{1}{l}
   \sum_{r}
   e^{i(k'-k)r}
   \sum_{r'} 
   \langle rm|r' n\rangle  
   e^{ik'(r'-r)}\\
   &=& 
   \delta_{kk'}S_{mn}(k),
\end{eqnarray}
we obtain
\begin{eqnarray}
\label{eq:eqn5-15}
  \langle rm \vert RM \rangle 
%  \approx
  = 
  \sum_{k} \frac{e^{ik (r-R-r_0 (M))}}{l} S_{mm' (M)}(k). 
\end{eqnarray}
We are now ready to evaluate $A_{kj,kj}$ given by Eq.~(\ref{eq:eqn5-6}). 
By using Eqs.~(\ref{eq:eqn5-7})-(\ref{eq:eqn5-10}), and (\ref{eq:eqn5-15}), the spectral function 
$A_{kj,kj}$ can be rewritten as
\begin{eqnarray}
\label{eq:eqn5-16}
 \nonumber
 \lefteqn{A_{kj,kj}
  = }\\
  &&
 \nonumber
  \sum_{KMNr' RR'} \frac{e^{-ik(R+r_0(M)-r')}}{lL} e^{iK(R-R')}\\
 && \times 
  C^{KJ}_{M} {C^{KJ}_{N}}^{*} \langle R' N|r' m' (M)\rangle A_{KJ,KJ}.
\end{eqnarray}
It should be noted in the derivation that $S_{nm}^{-1}(k)$ in Eq.~(\ref{eq:eqn5-6}) 
and $S_{mm' (M)}(k)$ in Eq.~(\ref{eq:eqn5-15}) are replaced by a relation 
$\sum_{m}S_{nm}^{-1}(k)S_{mm'(M)}(k)=\delta_{nm'(M)}$. 
The replacement is a crucial step in our derivation because Eq.~(\ref{eq:eqn5-16}) 
does not require any information about the position of the LCAO basis functions 
defined in the conceptual system, which is the reason why we call the {\it conceptual} 
system. The formula of Eq.~(\ref{eq:eqn5-16}) still has room to be simplified. 
By noting that 
\begin{eqnarray}
\label{eq:eqn5-17}
  \frac{1}{L} \sum_{R} e^{i(K-k)R} = \delta_{k-G,K},
\end{eqnarray}
and 
\begin{eqnarray}
\label{eq:eqn5-18}
 \nonumber
 \lefteqn{
 \sum_{r' R'} e^{-iKR'}e^{ikr'} \langle R' N|r' m'(M)\rangle 
  }\\
 \nonumber
  && = 
 \sum_{r' R'} e^{-i(k-G)R'}e^{ikr'} \langle R' N|r' m'(M)\rangle \\
 \nonumber
  && =  
 \sum_{R'} e^{iGR'} \sum_{r'} e^{ik(r'-R')} \langle 0 N|(r'-R') m'(M)\rangle \\
  && = 
 L \sum_{r'} e^{ikr'} \langle 0 N|r' m'(M)\rangle, 
\end{eqnarray}
we finally obtain 
\begin{eqnarray}
\label{eq:eqn5-19}
 A_{kj,kj}(\omega) = \frac{L}{l} \sum_{K} \delta_{k-G,K} W_{KJ} A_{KJ,KJ}(\omega)
\end{eqnarray}
with 
\begin{eqnarray}
\label{eq:eqn5-20}
 W_{KJ} = \sum_{MNr} e^{ik(r-r'(M))} C^{KJ}_{M} {C^{KJ}_{N}}^{*} S_{0N,rm(M)},  
\end{eqnarray}
where $A_{KJ,KJ}(\omega)$ is just a delta function at the eigenvalue, $\delta (\omega - \epsilon_{KJ})$.
In Eq.~(\ref{eq:eqn5-20}) the overlap integral $S_{0N,rm(M)}$ is evaluated by assuming 
the positions of LCAO basis functions used in the actual supercell calculation, which is 
one of our approximations that we made before. 
With the two approximations, one can calculate the spectral weight in terms of the normal cell 
representation without relying on the substance of the conceptual system.
Physically, the use of the conceptual system is just to have a guidance in obtaining the proper spectral 
weight in the desired representation. As can be confirmed, the sum of weight of the unfolded counterparts 
for each supercell eigenstate is exactly one.

It is also noted from Eq.~(\ref{eq:eqn5-17}) that the spectral function at $k$ is contributed 
from only the folded counterpart $A_{(k-G)J,(k-G)J}$. The formula of Eq.~(\ref{eq:eqn5-19})
allows us to separately calculate the unfolded spectral weight of each supercell eigenstate
within a given energy window by using the eigenvalue, the LCAO coefficients, and the overlap matrix 
elements obtained from the supercell calculation.
It becomes clear that the phase factor $e^{ik(r-r'(M))}$ plays an important role to determine 
a spectral weight of the extended-zone representation in unfolding the band structure of the supercell, 
and that the effect of translational symmetry breaking has been recorded in not only the LCAO 
coefficients, but also the overlap matrix elements in the supercell calculation. 
The formula is general for the localized orbitals and can be applied for both orthogonal and 
non-orthogonal basis functions. If Eq.~(\ref{eq:eqn5-19}) is applied for a supercell without any 
symmetry breakers, the band dispersion of the chosen normal cell is recovered in a precise 
mathematical sense. 

In addition, it is very interesting to see how each basis function contributes to the unfolded 
spectral weight. The decomposition to each contribution may be possible by defining $W_{KJM}^k$
as follows:
\begin{eqnarray}
\label{eq:eqn5-21}
 \nonumber
 W_{KJM}^k = \frac{L}{l} \delta_{k-G,K} C^{KJ}_{M}\sum_{rN}e^{ik(r-r'(M))}{C^{KJ}_{N}}^{*}S_{0N,rm(M)}.\\
\end{eqnarray}
Then, the spectral function can be written as 
\begin{eqnarray}
\label{eq:eqn5-22}
 A_{kj,kj}(\omega) = \sum_{K} A_{KJ,KJ}(\omega) \sum_{M} W_{KJM}^k.
\end{eqnarray}
By decomposing the spectral weight with Eq.~(\ref{eq:eqn5-21}) into the LCAO basis functions, one can 
analyze which localized basis functions compose each band dispersion, and which bands correspond
to spatially localized states such as surface states of a slab and vacancy states in a bulk.

Finally, we summarize the practical procedure to unfold the band structure obtained by the supercell 
calculation. First, one needs to introduce a supercell and a normal cell, and define a rule of the 
relabeling for $r'(M)$ and $m(M)$ in Eq.~(\ref{eq:eqn5-20}). The second step is to perform the band 
structure calculation of the supercell, where the LCAO coefficients and the overlap matrix elements 
are obtained. The third step is to calculate the spectral weight by using Eq.~(\ref{eq:eqn5-20}) or (\ref{eq:eqn5-21})
with the LCAO coefficients and the overlap matrix elements calculated at the second step.

\section{Electronic state of zirconium diboride slab}

\begin{figure}[b]
\includegraphics*[width=0.60\columnwidth,angle=90]{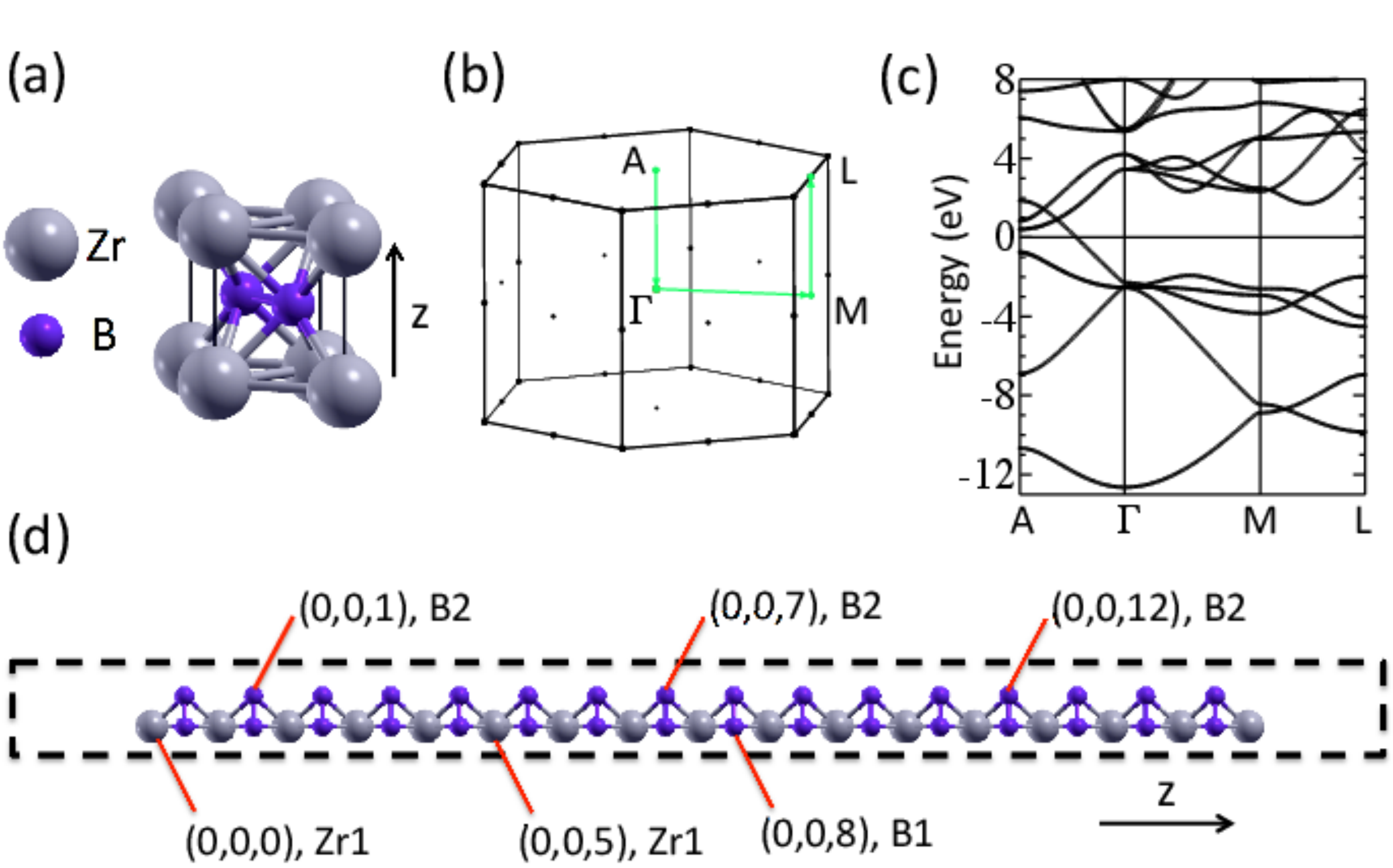}
\caption{\label{fig:fig2}
(a) The hexagonal unit cell of bulk ZrB$_2$ is presented with one Zr atom at (0,0,0) and two B atoms at (1/3,2/3,1/2) 
and (2/3,1/3,1/2). (b) The first BZ of the bulk ZrB$_2$. (c) The electronic band structure of bulk ZrB$_2$ along the 
path indicated in (b). (d) The calculated 33-layer Zr-terminated slab is shown with the (1$\times$1) unit cell 
of ZrB$_2$(0001) surface. The dashed rectangle gives an idea how thick the vacuum is. 
For the unfolding method, we choose the unit cell of bulk ZrB$_2$ as the normal cell.
Since only the difference between lattice vectors are needed in the unfolding formula,
we show one example for relabeling supercell atoms by the normal cell lattice vectors and atoms. 
The basis functions will follow the assignment of the corresponding atoms. 
}
\end{figure}

A common way to identify the surface state is to compare the band structure of the slab model with that of the bulk 
and/or to highlight the atomic contribution of the terminated layer. The idea can hardly work for a large unit cell 
of studied surfaces, where the heavily folded bands are involved. 
The large supercell is especially needed due to a spontaneous reconstruction\cite{Smeu} or the interaction 
with an added layer.\cite{Busse,Fleurence,vogt} 
The unfolding method can provide a way to the direct comparison of the band structures in the same BZ, 
which enables the visualization of the symmetry breaking and the surface states.
In the section, we demonstrate how our unfolding method can be applied to study the electronic states of a ZrB$_2$ slab, 
consisting of 17 Zr and 16 B layers, with the Zr-terminated (0001) surface, where the unit cell of bulk ZrB$_2$ is 
used as the normal unit cell.
Although the electronic states of Zr-terminated ZrB$_2$(0001) surface have been well studied,\cite{Aizawa,Kumashiro} 
we are motivated by a recent fabrication of silicene on the ZrB$_2$(0001) surface to reinvestigate 
the surface states in detail.\cite{Fleurence} 

The DFT calculations were performed by the OpenMX code which is based on norm-conserving pseudopotentials generated 
with multiple reference energies\cite{MBK} and linear combination of optimized pseudoatomic basis functions.\cite{Ozaki,openmx} 
The cutoff radius of 7.0 Bohr was used for all the basis functions. For each Zr atom, three, three, and two optimized 
radial functions were allocated for the $s$-, $p$-, and $d$-orbitals, respectively, as denoted by Zr-$s3p3d2$, 
while B-$s4p2d1$ were allocated for each B atom. The regular mesh of 270 Ry in real space was used for the numerical 
integrations and solution of the Poisson equation.\cite{Soler}
The exchange-correlation energy functional was treated with a generalized gradient approximation (GGA) proposed 
by Perdew, Burke and, Ernzerhof.\cite{Perdew} 
Geometrical structures were optimized until the maximum force on atom becomes less than $3\times10^{-4}$ Hartree/Bohr. 
In order to validate the accuracy of the DFT methods for the system, we optimized the lattice constants
of the bulk ZrB$_2$, where the primitive cell of the bulk ZrB$_2$ is hexagonal with one Zr atom at (0,0,0) 
and two B atoms at (1/3,2/3,1/2) and (2/3,1/3,1/2). The optimized lattice constants of bulk ZrB$_2$ were 
found to be $a=3.174$~\AA~and $c=3.550$~\AA~with a $k$-point sampling of 8$\times$8$\times$5 mesh, 
which are in good agreement with experimental data ($a=3.170$~\AA~and $c=3.533$~\AA) taken from Ref.~\onlinecite{Vajeeston}.
The geometrical structure and the band structure of the bulk ZrB$_2$ are shown in Fig.~\ref{fig:fig2}.

As a next step, we performed geometry optimization of the ZrB$_2$ slab with a $k$-point sampling of 
8$\times$8$\times$1 mesh, where the optimized lattice constants obtained by the bulk calculation were 
used for the $a$- and $b$-axes, and the vacuum of 14.292~\AA~ were inserted to avoid the interaction between the slabs. 
Since the unfolding formula can be applied to the calculation with or without adding basis functions in the vacant region,
no additional basis functions are added into the vacuum in this study and one can think the corresponding LCAO coefficients 
and the overlap matrix elements are zero.
The optimized structure exhibits that there is no surface reconstruction for the ZrB$_2$(0001) surface that 
has been checked with the (2$\times$2) unit cell of ZrB$_2$(0001) surface, and 
the outermost Zr-Zr distance is 3.492~\AA~which is slightly shorter than the bulk value (3.550~\AA). 
Up to this point, the overlap matrix elements between LCAO basis functions had already been obtained. 
In order to study the electronic band structure in the bulk representation, we need to employ the unfolding method
by performing the band structure calculation along the path shown in Fig.~\ref{fig:fig2} (b) where the LCAO coefficients 
are obtained. An additional effort in evaluating Eq.~(\ref{eq:eqn5-20}) or (\ref{eq:eqn5-21}) is to 
assign the supercell basis functions to the normal cell units. One example is provided for the
relabelings as shown in Fig.~\ref{fig:fig2} (d). Note that only the difference between lattice vectors are needed in the implementation.

Before showing the unfolded band structure, we remark several differences of our unfolding method in 
understanding the surface state from the conventional study which shows only the atomic weight of the folded band structure. 
First, the unfolded spectral function itself is physically meaningful and the developed broadening is related to the quasiparticle lifetime 
of an electron in the unfolded $\ket{kj}$ state.\cite{Faulkner,Onida} Our method not only shows the total weight
but also allows a detailed analysis of the contribution of each atomic basis function to the {\it unfolded} spectral function.
Since a thicker vacuum would break the translational symmetry of the bulk more, the unfolded weight reflects the 
relative lifetime for each reference $\ket{kj}$ electron that can survive in the slab system.
Secondly, even without the surface reconstruction, the folded bands due to the slab thickness still exist, which could result 
in many shadow bands and/or prominent splittings in the original bulk bands. The unfolding method recovers the 
proper spectral weight and make the visualization of the effect of symmetry breaking straightforwardly. One more interesting 
advantage is the capability of recovering the dispersion along the out-of-plane direction which is almost impossible for the 
conventional electronic band structure calculation to reveal such a dispersion. Recall that only one $k$-point sampling is 
needed along that direction in giving rise to horizontal bands.

\begin{figure}[t]
\includegraphics*[width=1.00\columnwidth,angle=0]{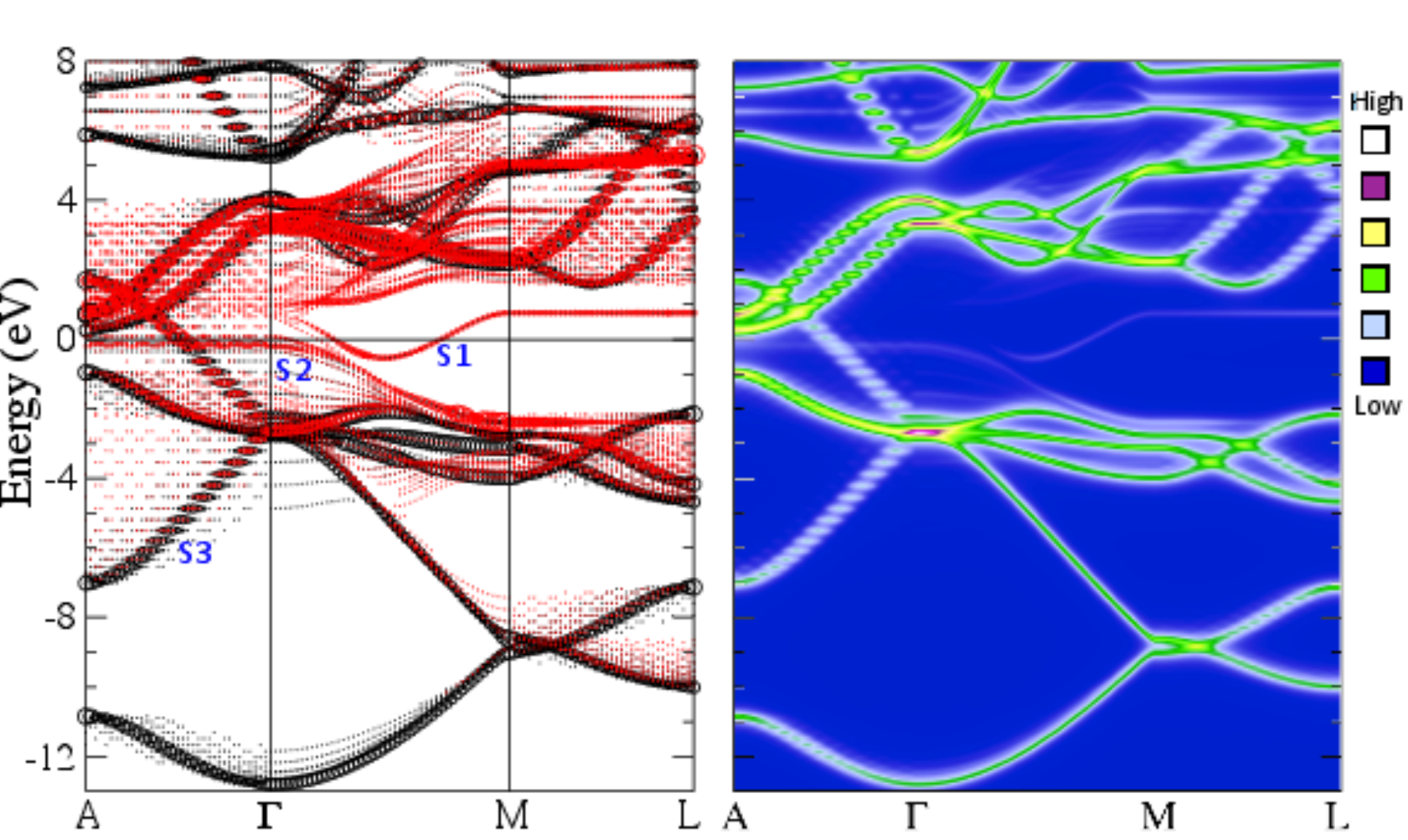}
\caption{\label{fig:fig3}
The left panel presents the unfolded band structure in the first BZ of bulk ZrB$_2$. The circle's radius is proportional 
to the calculated spectral weight for each $\ket{kj}$ basis. The outermost Zr $d$ contribution is colored red
and the sum of all Zr and B orbital contributions is colored black. The radii of the red circles are multiplied by ten
for a better visualization. The right panel shows the sum of all the unfolded total weight by a contour plot. The bands 
labeled by S1 and S2 can be easily identified as surface bands. The originally horizontal bands along the out-of-plane 
directions show a dispersive behavior via unfolding and are found to follow the bulk dispersion, for example, the band 
labeled by S3.
}
\end{figure}

We now present the unfolded band structure in Fig.~\ref{fig:fig3}. The result clearly shows that the overall bulk
band structure is well maintained, accompanied with some shadow bands and several additional interesting features. Without calculating
the proper spectral weight, it is difficult to identify the shadow band since each supercell eigenstate presents an equal amount of 
spectral weight. One interesting feature is the newly revealed surface bands in comparison to the bulk bands 
[c.f. Fig.~\ref{fig:fig2} (c)], for example, the bands labeled by S1 and S2 which consist of large contribution of the outermost Zr $d$ orbital. 
The electrons of the S1 band show diminishing spectral weight while the path approaches to the $\Gamma$ point, indicating negligible 
quasiparticle lifetimes and their short mean free paths. Note that the outermost atomic contribution would show a 
diminishing behavior while approaching to the $\Gamma$ point for the S1 band without unfolding.\cite{Aizawa}
However, our unfolding result demonstrates that the negligible spectral weight is expected even with the consideration of the contribution of 
all the layers.

For the dispersion in the out-of-plane direction, horizontal bands stemmed from the $\Gamma$- and $M$-point calculations 
are found, and can be identified as the slab states which are the quantized states due to electron confinement caused by the presence 
of the surface.\cite{Schulte,Chou} Our unfolding result shows that the discrete energy levels reveal a dispersive behavior as found in Fig.~\ref{fig:fig3}. 
The feature of dispersive quantized spectral weight can be understood in such a way that the thickness of the 
slab must construct the bulk property to some degrees. As a result, both the bulk dispersion and the quantized levels are observed
simultaneously. One can also see strong spectral broadenings in the 
out-of-plane direction from the center of the reference bulk spectral weight [c.f. Fig.~\ref{fig:fig2} (c)]. This shorter quasiparticle 
lifetime at large deviation from the bulk band is expected, which is missing in the conventional supercell band structure calculation. 
Of course, the lifetime is expected to be much longer at the bulk correspondence for a thicker slab, where the slab eigenstate is closer to 
the bulk eigenstate. All the interesting findings shown in Fig.~\ref{fig:fig3} are easily obtained by the unfolding method. 
Note that the quantum confinement effect has already been observed in multilayer graphene by ARPES.~\cite{Ohta}

\section{Summary}

By making use of a conceptual normal cell, we have derived a general formula in unfolding first-principles band structures, 
which can be applied to the LCAO basis functions being an ideal basis set for the unfolding concept. The translational symmetry breaking 
is recorded in the LCAO coefficients of the supercell eigenstates and built in the real-space locations of the basis functions. 
Once a conceptual normal cell is defined, the unfolded spectral weight is properly delivered via the phases of differences between normal cell
lattice vectors, the LCAO coefficients, and the overlap integrals between basis functions. 
The key effort for the unfolding procedure is just to define the conceptual normal cell in real space.
We have applied the method for studying the 
electronic states of ZrB$_2$(0001) slab along both the in-plane and the out-of-plane directions. A dispersive quantized 
spectral weight of the slab states with strong spectral broadenings in the out-of-plane direction can be easily revealed via unfolding. 
This interesting behavior should be found in all slabs in general, especially the thinner ones with layered structures. 
Since the LCAO basis functions are efficient in computation and the obtained supercell LCAO coefficients can be directly used in unfolding
the band structure, various applications including large-scale systems can be expected by using our unfolding method for studying 
the translational symmetry breaking.

\bibliography{refs}

\begin{thebibliography}{39}
\expandafter\ifx\csname natexlab\endcsname\relax\def\natexlab#1{#1}\fi
\expandafter\ifx\csname bibnamefont\endcsname\relax
  \def\bibnamefont#1{#1}\fi
\expandafter\ifx\csname bibfnamefont\endcsname\relax
  \def\bibfnamefont#1{#1}\fi
\expandafter\ifx\csname citenamefont\endcsname\relax
  \def\citenamefont#1{#1}\fi
\expandafter\ifx\csname url\endcsname\relax
  \def\url#1{\texttt{#1}}\fi
\expandafter\ifx\csname urlprefix\endcsname\relax\def\urlprefix{URL }\fi
\providecommand{\bibinfo}[2]{#2}
\providecommand{\eprint}[2][]{\url{#2}}

\bibitem[{\citenamefont{Kohn and Sham}(1965)}]{Kohn}
\bibinfo{author}{\bibfnamefont{W.}~\bibnamefont{Kohn}} \bibnamefont{and}
  \bibinfo{author}{\bibfnamefont{L.~J.} \bibnamefont{Sham}},
  \bibinfo{journal}{Phys. Rev.} \textbf{\bibinfo{volume}{140}},
  \bibinfo{pages}{A1133} (\bibinfo{year}{1965}).

\bibitem[{\citenamefont{Ashcroft and Mermin}(1976)}]{Ashcroft}
\bibinfo{author}{\bibfnamefont{N.~W.} \bibnamefont{Ashcroft}} \bibnamefont{and}
  \bibinfo{author}{\bibfnamefont{N.~D.} \bibnamefont{Mermin}},
  \emph{\bibinfo{title}{Solid State Physics}} (\bibinfo{publisher}{Holt,
  Rinehart, and Winston, New York}, \bibinfo{year}{1976}).

\bibitem[{\citenamefont{Martin}(2004)}]{Martin}
\bibinfo{author}{\bibfnamefont{R.~M.} \bibnamefont{Martin}},
  \emph{\bibinfo{title}{Electronic Structure: Basic Theory and Practical
  Methods}} (\bibinfo{publisher}{Cambridge University Press, New York},
  \bibinfo{year}{2004}).

\bibitem[{\citenamefont{Payne et~al.}(1992)\citenamefont{Payne, Teter, Allan,
  Arias, and Joannopoulos}}]{Payne}
\bibinfo{author}{\bibfnamefont{M.~C.} \bibnamefont{Payne}},
  \bibinfo{author}{\bibfnamefont{M.~P.} \bibnamefont{Teter}},
  \bibinfo{author}{\bibfnamefont{D.~C.} \bibnamefont{Allan}},
  \bibinfo{author}{\bibfnamefont{T.~A.} \bibnamefont{Arias}}, \bibnamefont{and}
  \bibinfo{author}{\bibfnamefont{J.~D.} \bibnamefont{Joannopoulos}},
  \bibinfo{journal}{Rev. Mod. Phys.} \textbf{\bibinfo{volume}{64}},
  \bibinfo{pages}{1045} (\bibinfo{year}{1992}).

\bibitem[{\citenamefont{Faulkner and Stocks}(1980)}]{Faulkner}
\bibinfo{author}{\bibfnamefont{J.~S.} \bibnamefont{Faulkner}} \bibnamefont{and}
  \bibinfo{author}{\bibfnamefont{G.~M.} \bibnamefont{Stocks}},
  \bibinfo{journal}{Phys. Rev. B} \textbf{\bibinfo{volume}{21}},
  \bibinfo{pages}{3222} (\bibinfo{year}{1980}).

\bibitem[{\citenamefont{Onida et~al.}(2002)\citenamefont{Onida, Reining, and
  Rubio}}]{Onida}
\bibinfo{author}{\bibfnamefont{G.}~\bibnamefont{Onida}},
  \bibinfo{author}{\bibfnamefont{L.}~\bibnamefont{Reining}}, \bibnamefont{and}
  \bibinfo{author}{\bibfnamefont{A.}~\bibnamefont{Rubio}},
  \bibinfo{journal}{Rev. Mod. Phys.} \textbf{\bibinfo{volume}{74}},
  \bibinfo{pages}{601} (\bibinfo{year}{2002}).

\bibitem[{\citenamefont{Lee et~al.}(2005)\citenamefont{Lee, Nardelli, and
  Marzari}}]{Marzari}
\bibinfo{author}{\bibfnamefont{Y.-S.} \bibnamefont{Lee}},
  \bibinfo{author}{\bibfnamefont{M.~B.} \bibnamefont{Nardelli}},
  \bibnamefont{and} \bibinfo{author}{\bibfnamefont{N.}~\bibnamefont{Marzari}},
  \bibinfo{journal}{Phys. Rev. Lett.} \textbf{\bibinfo{volume}{95}},
  \bibinfo{pages}{076804} (\bibinfo{year}{2005}).

\bibitem[{\citenamefont{Boykin and Klimeck}(2005)}]{Boykin}
\bibinfo{author}{\bibfnamefont{T.~B.} \bibnamefont{Boykin}} \bibnamefont{and}
  \bibinfo{author}{\bibfnamefont{G.}~\bibnamefont{Klimeck}},
  \bibinfo{journal}{Phys. Rev. B} \textbf{\bibinfo{volume}{71}},
  \bibinfo{pages}{115215} (\bibinfo{year}{2005}).

\bibitem[{\citenamefont{Popescu and Zunger}(2010)}]{Voicu}
\bibinfo{author}{\bibfnamefont{V.}~\bibnamefont{Popescu}} \bibnamefont{and}
  \bibinfo{author}{\bibfnamefont{A.}~\bibnamefont{Zunger}},
  \bibinfo{journal}{Phys. Rev. Lett.} \textbf{\bibinfo{volume}{104}},
  \bibinfo{pages}{236403} (\bibinfo{year}{2010}).

\bibitem[{\citenamefont{Popescu and Zunger}(2012)}]{Zunger}
\bibinfo{author}{\bibfnamefont{V.}~\bibnamefont{Popescu}} \bibnamefont{and}
  \bibinfo{author}{\bibfnamefont{A.}~\bibnamefont{Zunger}},
  \bibinfo{journal}{Phys. Rev. B} \textbf{\bibinfo{volume}{85}},
  \bibinfo{pages}{085201} (\bibinfo{year}{2012}).

\bibitem[{\citenamefont{Ku et~al.}(2010)\citenamefont{Ku, Berlijn, and
  Lee}}]{Wei}
\bibinfo{author}{\bibfnamefont{W.}~\bibnamefont{Ku}},
  \bibinfo{author}{\bibfnamefont{T.}~\bibnamefont{Berlijn}}, \bibnamefont{and}
  \bibinfo{author}{\bibfnamefont{C.-C.} \bibnamefont{Lee}},
  \bibinfo{journal}{Phys. Rev. Lett.} \textbf{\bibinfo{volume}{104}},
  \bibinfo{pages}{216401} (\bibinfo{year}{2010}).

\bibitem[{\citenamefont{Berlijn et~al.}(2011)\citenamefont{Berlijn, Volja, and
  Ku}}]{Tom}
\bibinfo{author}{\bibfnamefont{T.}~\bibnamefont{Berlijn}},
  \bibinfo{author}{\bibfnamefont{D.}~\bibnamefont{Volja}}, \bibnamefont{and}
  \bibinfo{author}{\bibfnamefont{W.}~\bibnamefont{Ku}}, \bibinfo{journal}{Phys.
  Rev. Lett.} \textbf{\bibinfo{volume}{106}}, \bibinfo{pages}{077005}
  (\bibinfo{year}{2011}).

\bibitem[{\citenamefont{Konbu et~al.}(2011)\citenamefont{Konbu, Nakamura,
  Ikeda, and Arita}}]{Arita}
\bibinfo{author}{\bibfnamefont{S.}~\bibnamefont{Konbu}},
  \bibinfo{author}{\bibfnamefont{K.}~\bibnamefont{Nakamura}},
  \bibinfo{author}{\bibfnamefont{H.}~\bibnamefont{Ikeda}}, \bibnamefont{and}
  \bibinfo{author}{\bibfnamefont{R.}~\bibnamefont{Arita}}, \bibinfo{journal}{J.
  Phys. Soc. Jpn.} \textbf{\bibinfo{volume}{80}}, \bibinfo{pages}{123701}
  (\bibinfo{year}{2011}).

\bibitem[{\citenamefont{Kang et~al.}(2012)}]{Kang}
\bibinfo{author}{\bibfnamefont{J.-S.} \bibnamefont{Kang}} \bibnamefont{et~al.},
  \bibinfo{journal}{Phys. Rev. B} \textbf{\bibinfo{volume}{85}},
  \bibinfo{pages}{085104} (\bibinfo{year}{2012}).

\bibitem[{\citenamefont{Marzari and Vanderbilt}(1997)}]{Vanderbilt}
\bibinfo{author}{\bibfnamefont{N.}~\bibnamefont{Marzari}} \bibnamefont{and}
  \bibinfo{author}{\bibfnamefont{D.}~\bibnamefont{Vanderbilt}},
  \bibinfo{journal}{Phys. Rev. B} \textbf{\bibinfo{volume}{56}},
  \bibinfo{pages}{12847} (\bibinfo{year}{1997}).

\bibitem[{\citenamefont{Ku et~al.}(2002)\citenamefont{Ku, Rosner, Pickett, and
  Scalettar}}]{Scalettar}
\bibinfo{author}{\bibfnamefont{W.}~\bibnamefont{Ku}},
  \bibinfo{author}{\bibfnamefont{H.}~\bibnamefont{Rosner}},
  \bibinfo{author}{\bibfnamefont{W.~E.} \bibnamefont{Pickett}},
  \bibnamefont{and} \bibinfo{author}{\bibfnamefont{R.~T.}
  \bibnamefont{Scalettar}}, \bibinfo{journal}{Phys. Rev. Lett.}
  \textbf{\bibinfo{volume}{89}}, \bibinfo{pages}{167204}
  (\bibinfo{year}{2002}).

\bibitem[{\citenamefont{van Heumen et~al.}(2011)}]{Koepernik}
\bibinfo{author}{\bibfnamefont{E.}~\bibnamefont{van Heumen}}
  \bibnamefont{et~al.}, \bibinfo{journal}{Phys. Rev. Lett.}
  \textbf{\bibinfo{volume}{106}}, \bibinfo{pages}{027002}
  (\bibinfo{year}{2011}).

\bibitem[{\citenamefont{Blum et~al.}(2009)}]{Blum}
\bibinfo{author}{\bibfnamefont{V.}~\bibnamefont{Blum}} \bibnamefont{et~al.},
  \bibinfo{journal}{Comput. Phys. Commun.} \textbf{\bibinfo{volume}{180}},
  \bibinfo{pages}{2175} (\bibinfo{year}{2009}).

\bibitem[{\citenamefont{Soler et~al.}(2002)}]{Soler}
\bibinfo{author}{\bibfnamefont{J.~M.} \bibnamefont{Soler}}
  \bibnamefont{et~al.}, \bibinfo{journal}{J. Phys. Condens. Matter}
  \textbf{\bibinfo{volume}{14}}, \bibinfo{pages}{2745} (\bibinfo{year}{2002}).

\bibitem[{\citenamefont{Frisch et~al.}()}]{Frisch}
\bibinfo{author}{\bibfnamefont{M.~J.} \bibnamefont{Frisch}}
  \bibnamefont{et~al.}, \emph{\bibinfo{title}{Gaussian~09 {R}evision {A}.1}},
  \bibinfo{note}{gaussian Inc. Wallingford CT 2009}.

\bibitem[{\citenamefont{Andersen}(1975)}]{Andersen}
\bibinfo{author}{\bibfnamefont{O.~K.} \bibnamefont{Andersen}},
  \bibinfo{journal}{Phys. Rev. B} \textbf{\bibinfo{volume}{12}},
  \bibinfo{pages}{3060} (\bibinfo{year}{1975}).

\bibitem[{\citenamefont{Ozaki}(2003)}]{Ozaki}
\bibinfo{author}{\bibfnamefont{T.}~\bibnamefont{Ozaki}},
  \bibinfo{journal}{Phys. Rev. B} \textbf{\bibinfo{volume}{67}},
  \bibinfo{pages}{155108} (\bibinfo{year}{2003}).

\bibitem[{\citenamefont{Yang}(1991)}]{Yang}
\bibinfo{author}{\bibfnamefont{W.}~\bibnamefont{Yang}}, \bibinfo{journal}{Phys.
  Rev. Lett.} \textbf{\bibinfo{volume}{66}}, \bibinfo{pages}{1438}
  (\bibinfo{year}{1991}).

\bibitem[{\citenamefont{Ozaki}(2006)}]{Ozaki2}
\bibinfo{author}{\bibfnamefont{T.}~\bibnamefont{Ozaki}},
  \bibinfo{journal}{Phys. Rev. B} \textbf{\bibinfo{volume}{74}},
  \bibinfo{pages}{245101} (\bibinfo{year}{2006}).

\bibitem[{\citenamefont{Ozaki}(2010)}]{Ozaki3}
\bibinfo{author}{\bibfnamefont{T.}~\bibnamefont{Ozaki}},
  \bibinfo{journal}{Phys. Rev. B} \textbf{\bibinfo{volume}{82}},
  \bibinfo{pages}{075131} (\bibinfo{year}{2010}).

\bibitem[{\citenamefont{Fleurence et~al.}(2012)}]{Fleurence}
\bibinfo{author}{\bibfnamefont{A.}~\bibnamefont{Fleurence}}
  \bibnamefont{et~al.}, \bibinfo{journal}{Phys. Rev. Lett.}
  \textbf{\bibinfo{volume}{108}}, \bibinfo{pages}{245501}
  (\bibinfo{year}{2012}).

\bibitem[{\citenamefont{Ohta et~al.}(2007)\citenamefont{Ohta, Bostwick,
  McChesney, Seyller, Horn, and Rotenberg}}]{Ohta}
\bibinfo{author}{\bibfnamefont{T.}~\bibnamefont{Ohta}},
  \bibinfo{author}{\bibfnamefont{A.}~\bibnamefont{Bostwick}},
  \bibinfo{author}{\bibfnamefont{J.~L.} \bibnamefont{McChesney}},
  \bibinfo{author}{\bibfnamefont{T.}~\bibnamefont{Seyller}},
  \bibinfo{author}{\bibfnamefont{K.}~\bibnamefont{Horn}}, \bibnamefont{and}
  \bibinfo{author}{\bibfnamefont{E.}~\bibnamefont{Rotenberg}},
  \bibinfo{journal}{Phys. Rev. Lett.} \textbf{\bibinfo{volume}{98}},
  \bibinfo{pages}{206802} (\bibinfo{year}{2007}).

\bibitem[{\citenamefont{Neto et~al.}(2009)\citenamefont{Neto, Guinea, Peres,
  Novoselov, and Geim}}]{Castro}
\bibinfo{author}{\bibfnamefont{A.~H.~C.} \bibnamefont{Neto}},
  \bibinfo{author}{\bibfnamefont{F.}~\bibnamefont{Guinea}},
  \bibinfo{author}{\bibfnamefont{N.~M.~R.} \bibnamefont{Peres}},
  \bibinfo{author}{\bibfnamefont{K.~S.} \bibnamefont{Novoselov}},
  \bibnamefont{and} \bibinfo{author}{\bibfnamefont{A.~K.} \bibnamefont{Geim}},
  \bibinfo{journal}{Rev. Mod. Phys} \textbf{\bibinfo{volume}{81}},
  \bibinfo{pages}{109} (\bibinfo{year}{2009}).

\bibitem[{\citenamefont{Vogt et~al.}(2012)}]{vogt}
\bibinfo{author}{\bibfnamefont{P.}~\bibnamefont{Vogt}} \bibnamefont{et~al.},
  \bibinfo{journal}{Phys. Rev. Lett.} \textbf{\bibinfo{volume}{108}},
  \bibinfo{pages}{155501} (\bibinfo{year}{2012}).

\bibitem[{\citenamefont{Smeu et~al.}(2012)}]{Smeu}
\bibinfo{author}{\bibfnamefont{M.}~\bibnamefont{Smeu}} \bibnamefont{et~al.},
  \bibinfo{journal}{Phys. Rev. B} \textbf{\bibinfo{volume}{85}},
  \bibinfo{pages}{195315} (\bibinfo{year}{2012}).

\bibitem[{\citenamefont{Busse et~al.}(2011)}]{Busse}
\bibinfo{author}{\bibfnamefont{C.}~\bibnamefont{Busse}} \bibnamefont{et~al.},
  \bibinfo{journal}{Phys. Rev. Lett.} \textbf{\bibinfo{volume}{107}},
  \bibinfo{pages}{036101} (\bibinfo{year}{2011}).

\bibitem[{\citenamefont{Aizawa et~al.}(2005)}]{Aizawa}
\bibinfo{author}{\bibfnamefont{T.}~\bibnamefont{Aizawa}} \bibnamefont{et~al.},
  \bibinfo{journal}{Phys. Rev. B} \textbf{\bibinfo{volume}{71}},
  \bibinfo{pages}{165405} (\bibinfo{year}{2005}).

\bibitem[{\citenamefont{Kumashiro et~al.}(2006)}]{Kumashiro}
\bibinfo{author}{\bibfnamefont{S.}~\bibnamefont{Kumashiro}}
  \bibnamefont{et~al.}, \bibinfo{journal}{e-J. Surf. Sci. Nanotech.}
  \textbf{\bibinfo{volume}{4}}, \bibinfo{pages}{100} (\bibinfo{year}{2006}).

\bibitem[{\citenamefont{Morrison et~al.}(1993)\citenamefont{Morrison, Bylander,
  and Kleinman}}]{MBK}
\bibinfo{author}{\bibfnamefont{I.}~\bibnamefont{Morrison}},
  \bibinfo{author}{\bibfnamefont{D.}~\bibnamefont{Bylander}}, \bibnamefont{and}
  \bibinfo{author}{\bibfnamefont{L.}~\bibnamefont{Kleinman}},
  \bibinfo{journal}{Phys. Rev. B} \textbf{\bibinfo{volume}{47}},
  \bibinfo{pages}{6728} (\bibinfo{year}{1993}).

\bibitem[{\citenamefont{Ozaki et~al.}()}]{openmx}
\bibinfo{author}{\bibfnamefont{T.}~\bibnamefont{Ozaki}} \bibnamefont{et~al.},
  \urlprefix\url{http://www.openmx-square.org/}.

\bibitem[{\citenamefont{Perdew et~al.}(1996)\citenamefont{Perdew, Burke, and
  Ernzerhof}}]{Perdew}
\bibinfo{author}{\bibfnamefont{J.~P.} \bibnamefont{Perdew}},
  \bibinfo{author}{\bibfnamefont{K.}~\bibnamefont{Burke}}, \bibnamefont{and}
  \bibinfo{author}{\bibfnamefont{M.}~\bibnamefont{Ernzerhof}},
  \bibinfo{journal}{Phys. Rev. Lett.} \textbf{\bibinfo{volume}{77}},
  \bibinfo{pages}{3865} (\bibinfo{year}{1996}).

\bibitem[{\citenamefont{Vajeeston et~al.}(2001)\citenamefont{Vajeeston,
  Ravindran, and adn R.~Asokamani}}]{Vajeeston}
\bibinfo{author}{\bibfnamefont{P.}~\bibnamefont{Vajeeston}},
  \bibinfo{author}{\bibfnamefont{P.}~\bibnamefont{Ravindran}},
  \bibnamefont{and} \bibinfo{author}{\bibfnamefont{C.~R.} \bibnamefont{adn
  R.~Asokamani}}, \bibinfo{journal}{Phys. Rev. B}
  \textbf{\bibinfo{volume}{63}}, \bibinfo{pages}{045115}
  (\bibinfo{year}{2001}).

\bibitem[{\citenamefont{Schulte}(1976)}]{Schulte}
\bibinfo{author}{\bibfnamefont{F.~K.} \bibnamefont{Schulte}},
  \bibinfo{journal}{Surf. Sci} \textbf{\bibinfo{volume}{55}},
  \bibinfo{pages}{427} (\bibinfo{year}{1976}).

\bibitem[{\citenamefont{Wei and Chou}(2002)}]{Chou}
\bibinfo{author}{\bibfnamefont{C.~M.} \bibnamefont{Wei}} \bibnamefont{and}
  \bibinfo{author}{\bibfnamefont{M.~Y.} \bibnamefont{Chou}},
  \bibinfo{journal}{Phys. Rev. B} \textbf{\bibinfo{volume}{66}},
  \bibinfo{pages}{233408} (\bibinfo{year}{2002}).

\end{thebibliography}
\end{document}